\documentstyle[12pt]{article}
\title{The Global Minimum of Energy Is Not Always a Sum of Local Minima - a
Note
on Frustration}
\author{Jacek Mi\c{e}kisz \\ Institut de Physique Th\'{e}orique \\
Universit\'{e} Catholique de Louvain \\ Chemin du Cyclotron, 2 \\ B-1348
Louvain-la-Neuve, Belgium.}
\begin{document}
\baselineskip=26pt
\maketitle
{\bf Abstract.} A classical lattice gas model with translation-invariant
finite
range competing interactions, for which there does not exist an equivalent
translation-invariant finite range nonfrustrated potential, is constructed.
The
construction uses the structure of nonperiodic ground state configurations
of
the model. In fact, the model does not have any periodic ground state
configurations. However, its ground state - a translation-invariant
probability
measure supported by ground state configurations - is unique.


\eject

\section{Introduction}
Low temperature behavior of systems of many interacting particles results
from
the competition between energy and entropy, i.e., the minimization of the
free
energy. At zero temperature this reduces to the minimization of the
energy density. Configurations of a system which minimize its energy
density are
called ground state configurations. One of the important problems of
statistical
mechanics is to find ground state configurations for given interactions
between
particles. If we can find a configuration such that potential energies of
all
interactions between particles are minimal then we can conclude that it is
a
ground state configuration. It is then said that such a model is not
frustrated.
Otherwise, we may rearrange potentials and construct an equivalent
Hamiltonian
which may not be frustrated and which will enable us to find ground state
configurations. Here we present a classical lattice gas model with
translation-invariant finite range competing interactions for which there
does
not exist an equivalent translation-invariant finite range nonfrustrated
potential. In other words: the global minimum of energy is not the sum of
its
minima attained locally in space. More precisely, one cannot minimize the
energy density of
interacting particles by minimizing their energy in a finite box and all
its translates, no matter
how large is the box.

\section{Classical Lattice Gas Models, Frustration, and M-Potentials}

A classical latice gas model is a system in which every site of a lattice
$Z^{d}
$ can be occupied by one of $n$ different particles. An infinite lattice
configuration is an assignment of particles to lattice sites, that is an
element
of $\Omega=\{1,...,n\}^{Z^{d}}$. Particles can interact through many-body
potentials. A {\em potential} $\Phi$ is a collection of real valued
functions
$\Phi_{\Lambda}$ on configuration spaces
$\Omega_{\Lambda}=\{1,...,n\}^{\Lambda}$ for all finite $\Lambda \subset
Z^{d}$.
Here we assume $\Phi$ to have finite range, that is $\Phi_{\Lambda} = 0$ if
the
diameter of $\Lambda$ is large enough, and be translation-invariant. The
formal Hamiltonian can be
then written as $$H= \sum_{\Lambda}\Phi_{\Lambda}.$$
Two configurations $X,Y \in \Omega$ are said to be {\em equal at infinity},
$X
\sim Y$, if there exists a finite $\Lambda \subset Z^{d}$ such that $X = Y$
outside $\Lambda.$ The relative Hamiltonian is defined by
$$H(X,Y)=\sum_{\Lambda}(\Phi_{\Lambda}(X)-\Phi_{\Lambda}(Y)) \; \;for \; \;X
\sim
Y.$$ $X \in \Omega$ is a {\em ground state configuration} of $H$ if
$$H(Y,X) \geq 0 \; \; for \; \; any \; \; Y \sim X.$$ For any potential the
set
of ground state configurations is nonempty but it may not contain any
periodic
configurations \cite{rad1,rad2,ram}. We will be concerned here with
nonperiodic
ground state configurations which have uniformly defined frequencies for
all
finite patterns. By definition the orbit closure of such a ground state
configuration supports
a unique strictly ergodic translation-invariant measure called a ground
state
which is a zero temperature limit of a low temperature Gibbs state (an
infinite
volume grand canonical probability distribution). If we can find a
configuration that minimizes all $\Phi_{\Lambda}$, then it is necessarily a
ground state configuration and we call such potential nonfrustrated or an
m-potential \cite{slaw1,slaw2}. Formally, a potential $\Phi$ is an
{\em m-potential} if there exists a configuration $X$ such that

$$\Phi_{\Lambda}(X)=min_{Y}\Phi_{\Lambda}(Y) \; \; for \; \; any \; \;
finite \;
\; \Lambda.$$
Otherwise, we may try to rearrange interactions to obtain an equivalent
m-potential. Two potentials are defined to be {\em equivalent} if they
yield the
same relative Hamiltonian and therefore have the same ground state
configurations and the same Gibbs states. It follows that for any periodic
configuration its
energy density is the same (up to a trivial additive constant which can be
chosen to be zero) for all
equivalent potentials. It is best illustrated by an example of the
antiferromagnetic nearest neighbor
spin $1/2$ model on the triangular lattice. The formal Hamiltonian can be
written as follows:

$$H = \sum_{i,j}\sigma_{i}\sigma_{j},$$
where $\sigma_{i}, \sigma_{j}= \pm 1$ and $i$ and $j$ are nearest neighbor
sites on the triangular lattice. When you look at an elementary triangle it
is
easy to see that at least one pair of spins does not minimize its
interaction.
Two spins allign themselves in opposite directions and then the third one
can
minimize only one of the two remaining interactions. This choice is a
source of
frustration \cite{tou} (see also another approach to frustration
\cite{and,mie}). However,
we may construct the following equivalent potential:
$$\phi_{\triangle} =
1/2(\sigma_{i}\sigma_{j} + \sigma_{j}\sigma_{k} + \sigma_{k}\sigma_{i}),$$
where
$i$, $j$, and $k$ are vertices of an elementary triangle $\triangle$ and
$\Phi_{\Lambda}=0$ otherwise. Now, there are ground state configurations
minimizing every $\Phi_{\Lambda}.$ Three spins on every elementary triangle
still face choices but they act collectively and therefore are not
frustrated.

In the following section we construct an example of a lattice gas model
with nearest neighbor translation-invariant frustrated interactions for
which
there does not exist an equivalent finite range translation-invariant
m-potential. The main problem of proving the impossibility of an
m-potential is
that a grouping of interactions in big plaquettes, like in the above
example, is
not the only way of constructing an equivalent m-potential. To construct it
one may also
use an information about a global structure of excitations. In some models
just
grouping is clearly impossible because energy can be lower locally than
that of
a ground state configuration and you can pay for it arbitrarily far away,
yet
one can still construct an equivalent m-potential. One of the easiest
examples is a
one-dimensional Ising model with the following interactions: the energy
of $+-$ neighbors is equal to $-1$, the energy of $-+$ neighbors is $2$,
and
otherwise the energy is zero. There are arbitrarily long line segments with
the energy equal to $-1$.
Nevertheless, the above potential is equivalent to an m-potential with the
energy of $-+$ neighbors equal to $1$ and zero otherwise, or
$-1/4(\sigma_{i}\sigma_{i+1}-1)$ using
spin variables.

\section{An Intrinsically Frustrated Model}
The model is based on Robinson's
tiles \cite{rob,pat}. There is a family of 56 square-like tiles such that using
an infinite number of copies of each of them one can tile the plane only in
a
nonperiodic fashion. This can be translated into a lattice gas model in the
following way first introduced by Radin \cite{rad1,rad2,ram}. Every site of
the
square lattice can be occupied by one of the 56 different particles-tiles.
Two
nearest neighbor particles which do not ``match'' contribute positive
energy which we chose to be
$24$ for our purposes; otherwise, the energy is zero. Such a model
obviously does not have
periodic ground state configurations. There are uncountably many ground
state configurations but
only one translation-invariant ground state measure supported by them.
There is a one-to-one
correspondence between ground state configurations in the support of this
measure and Robinson's
nonperiodic tilings. Low temperature behavior of this model was
investigated in \cite{mie1,mie2,mie3}.

We describe now slightly modified Robinson tiles (with a different number
of tiles); we follow
\cite{rob} closely. There are seven basic tiles represented symbolically in
Fig.1. The rest of them
can be obtained by rotations and reflections. The first tile on the left is
called a cross; the rest
are called arms. All tiles are furnished with one of the four parity
markings shown in Fig.2.
The crosses can be combined with the parity marking at the lower left in
Fig.2. Vertical arms (the direction of long arrows) can be
combined with the marking at the lower right and horizontal arms with
the marking at the upper left. All tiles may be combined with the remaining
marking. Two nearest neighbor tiles ``match'' if arrow heads meet arrow
tails separately for
the parity markings and the markings of crosses and arms.

Let us now describe main features of Robinson's nonperiodic tilings.
We will concentrate on the lattice positions of crosses denoted by
$\lfloor,
\lceil, \rfloor, \rceil$, where directions of line segments correspond to
double
arrows in Fig.1. Every odd-odd position on the $Z^{2}$ lattice (if columns
and rows are suitably
numbered) is occupied by these tiles in relative orientations as in Fig.3.
They form the periodic
configuration with the period 4. Then in the center of each ``square'' one
has
to put again a cross such that the previous pattern reproduces but this
time
with the period 8. Continuing this procedure infinitely many times we
obtain a
nonperiodic configuration. It has built in periodic configurations of
period
$2^{n}$, $n\geq 2$ on sublattices of $Z^{2}$ as shown in Fig.4.

Now we will modify the above model introducing another level
of markings which are optional, that it is to say they can be present (one
at a time) or absent in
appropriate tiles.  Every cross can be equipped with one of the two
markings shown at the left in
Fig.5. The orientation of a marking at the top should be the same as the
orientation of its
cross and it comes in either red or yellow color; the second marking is
red. Arms can be furnished
with red or yellow lines shown in the middle column in Fig.5 (colored lines
should be parallel to
long arrows). Arms at the top in Fig.1 can be equipped with a marking at
the upper right in
Fig.5 with yellow-red, red-yellow, or yellow-yellow (but not red-red)
segments perpendicular to long
arrows and closer to their tails than to their heads or a red marking at
the lower right in Fig.5 with
a short segment parallel to long arrows and pointing in the other
direction. Finally, arms in the
middle row and at the bottom in Fig.1 can be equipped with a marking at the
upper right in Fig.5 with
yellow-red, red-yellow, yellow-yellow, and red-red segments perpendicular
to long arrows. Now, two
tiles match if in addition to previous requirements there are no broken
lines of new markings and
adjacent colors are the same. In the corresponding classical lattice gas
model in addition to
two-body nearest neighbor interactions we introduce a chemical potential
equal to $1$ for yellow
crosses and having a negative value $\tau > -1$ for red crosses with the
marking at the upper left in
Fig.5 (called simply red crosses from now on) and zero value for remaining
particles-tiles. Let
us
notice that in the absence of broken bonds our matching rules force the
number of yellow crosses to be
at least equal to the number of red crosses.

\newtheorem{ground}{Proposition}
\newtheorem{frus}{Theorem}
\begin{ground}
For $\tau >-1$ the unique ground state measure of the modified model is the
same
as of the original Robinson model.
\end{ground}
{\em Proof:} Let a broken bond be a unit segment on the dual lattice
separating two nearest neighbor
particles with a positive interaction energy (a common side of two nearest
neighbor tiles which do not match). Let us divide the lattice into
connected components without
broken bonds (two lattice sites are connected if they are nearest
neighbors) such that on every
component crosses having a relative orientation shown in Fig.3 form a
$2Z^{2}$ sublattice. This can
be achieved by paths on the lattice dual to $Z^{2}$ with lengths smaller
than $8$ and joining broken
bonds. For any such component either all crosses on its $2Z^{2}$ sublattice
are colored and then no
other cross can be colored or no cross on this sublattice is colored. This
follows from the fact that
there are four lines of colored arms emanating from each cross and such
lines cannot intersect each
other. We call a cross on a $2Z^{2}$ ($2^{n}Z^{2}, n \geq 1$, in general)
sublattice a boundary cross
if its distance from the boundary of its connected component is smaller
than $2$ ($2^{n}, n \geq
1$, in general).  Now, we decompose further every connected component
without colored
crosses on its $2Z^{2}$ sublattice into connected components such that
crosses form there a $4Z^{2}$
sublattice. This time it is achieved by paths on the dual lattice with
lengths smaller than $16$ and
again joining broken bonds. Again, for any such component either all
crosses on its $4Z^{2}$
sublattice are colored and then no other cross is colored or no cross on
this sublattice is colored.
In the latter case we have to decompose further this component. We repeat
this procedure for all
$2^{n}Z^{2}$ sublattices for every $n \geq 1$. Now, the total number of all
paths and every $n$ in
every finite region is bounded above by three times the number of broken
bonds (broken bonds and
paths can be regarded as vertices and edges of a planar graph and our bound
follows from Euler's
formula). This shows that the density of red boundary crosses is bounded
above by $24$ times the
density of broken bonds.The negative chemical potential of red boundary
crosses is then compensated by the
positive energy of broken bonds. On the other hand the density of yellow
crosses is at least equal to the
density of red crosses which are not interior ones so the negative chemical
potential of red interior
crosses is compensated by positive chemical potential of yellow crosses
$(\tau > -1)$. It follows
that the energy of a configuration is at least proportional to the total
length of broken bonds (a
Peierls condition is satisfied). Hence in any ground state configuration in
the support of a ground
state measure broken bonds are absent. Among configurations without
broken bonds, configurations without any colored particles (Robinsons's
original configurations) have
the minimal energy density (equal to zero) and are therefore the only
ground state
configurations. This follows again from the fact that $\tau > -1$ and the
number of yellow
crosses is greater or equal to the number of red crosses. $\Box$

Obviously, our interactions do not constitute an
m-potential. Moreover, it is impossible to construct a
translation-invariant
finite range m-potential by grouping interactions in big plaquettes like it
was done in the antiferromagnetic example. One may locate colored crosses
on
the sublattice $2^{n}Z^{2}$ therefore decreasing energy
locally and paying for it arbitrarily far away [see Fig.6]. Now we will
prove that for some
$\tau$  an equivalent translation-invariant finite range m-potential does
not
actually exist.
\begin{frus}
The above described model for $-1 < \tau < -6/10$ does not have an equivalent
translation-invariant finite range m-potential.
\end{frus}
{\em Proof:} Let us assume otherwise and let its range be smaller than
$2^{n}.$
Let us  consider three configurations with periodic arrangements of colored
markings and with their
periods shown in Fig.6, where squares have size $2^{n+1}$ and only colored
markings of the central
tile of each square are shown. Equating their energy densities for the
original interaction and a
hypothetical equivalent m-potential we obtain:
\begin{equation}
\tau+3=a_{r}+b_{ry}+c_{y}+d_{yr}+e_{ry}+g_{y}+i_{y}+j_{y}+k_{y}+l_{y}+m_{yr}
o_{y},
\end{equation}
\begin{equation}
2\tau+2=a_{r}+b_{ry}+c_{y}+d_{yr}+e_{r}+f_{r}+g_{y}+i_{r}+j_{ry}+k_{y}+l_{ry
+m_{r}+o_{y},
\end{equation}
\begin{equation}
2\tau+2=a_{r}+b_{r}+c_{r}+d_{r}+e_{ry}+f_{r}+g_{ry}+i_{y}+j_{y}+k_{y}+l_{y}+
_{yr}+o_{yr},
\end{equation}
where on the right hand sides we have nonnegative  contributions to
energy due to a hypothetical m-potential (we know from the Proposition that
the
nergy density of the ground state is zero so an m-potential cannot be
negative) and coming from
regions labelled in the  upper left corners of the squares in Fig.6;
subscripts correspond to
configurations of optional markings with $r$ denoting red, $y$ denoting
yellow, and $ry$, $yr$
meaning a change of colors along a line of arms.

Now, set $\tau =-1+\delta /2.$ From (2) we obtain $a_{r} \leq \delta$
and $b_{ry}+c_{y}+d_{yr}+g_{y}+o_{y} \leq \delta$, and from (3)
$e_{ry}+i_{y}+j_{y}+k_{y}+l_{y}+m_{yr} \leq \delta.$ Then it follows from
(1) that
$a_{r} \geq 2-(3/2)\delta$ which contradicts $a_{r} \leq \delta$ if
$\delta<4/5.$ This contradiction rules out the existence of an equivalent
translation-invariant finite range m-potential. $\Box$

\section{Conclusions}
A classical lattice gas model with
translation-invariant nearest neighbor competing interactions is
constructed.
Its unique translation-invariant ground state measure is supported by
nonperiodic ground state
configurations. There are local excitations in the model such that
the energy is locally lower than that of a ground state configuration and
one
pays for it arbitrarily far away. This shows that by grouping interactions
in big
plaquettes, like in the antiferromagnetic model on the triangular lattice,
one
cannot construct an equivalent finite range m-potential. More generally it
is proved that such a
potential actually does not exist. The model is therefore intrinsically
frustrated.

Let us note that in the antiferromagnetic model a spin on an
elementary triangle is frustrated because it faces a choice of direction.
Its both choices can be present in a ground state configuration making
therefore a ground state highly degenerate. In our example a particle may
choose a local minimum of energy  and then it appears that this does not
lead to
a ground state configuration.\\

\vspace{5mm}

{\bf Acknowledgments.} I
would like to thank Alan Sokal and Roberto Fernandez for an inspiration,
and
Jean Bricmont for helpful discussions. Bourse de recherche UCL/FDS is
gratefully acknowledged for the financial support.

\end{document}